\documentclass[final,5p,times,twocolumn]{elsarticle}

\usepackage{amssymb}
\usepackage{amsmath}
\usepackage{amssymb}
\usepackage{arydshln}
\usepackage{comment}

\usepackage[colorlinks=true, allcolors=blue]{hyperref}

\usepackage{lineno}
\usepackage[normalem]{ulem}

\begin{document}

\begin{frontmatter}



\title{Cylindrical Cavity Expansion: A Novel Method for Characterizing the Mechanical Properties of Soft Materials\\
}

\author[inst1,inst2,inst3]{Jian Li}
\author[inst1,inst2,inst3]{Zihao Xie}
\author[inst4]{Hannah Varner}
\author[inst5]{Chockalingam Senthilnathan}
\author[inst4,inst6]{Tal Cohen \corref{mycorrespondingauthor}}
\cortext[mycorrespondingauthor]{Corresponding author}
\ead{talco@mit.edu}

\affiliation[inst1]{organization={The State Key Laboratory of Heavy-duty and Express High-power Electric Locomotive},
            city={Changsha},
            postcode={410075}, 
            country={China}}

\affiliation[inst2]{organization={Key Laboratory of Trafﬁc Safety on Track of Ministry of Education, School of Traffic \& Transportation Engineering, Central South University},
            city={Changsha},
            postcode={410075}, 
            country={China}}

\affiliation[inst3]{organization={National \& Local Joint Engineering Research Center of Safety Technology for Rail Vehicle},
            city={Changsha},
            postcode={410075}, 
            country={China}}

\affiliation[inst4]{organization={Department of Mechanical Engineering, Massachusetts Institute of Technology}, 
            city={Cambridge},
            postcode={02139}, 
            state={MA},
            country={USA}}

\affiliation[inst5]{organization={Department of Aeronautics and Astronautics, Massachusetts Institute of Technology},
            city={Cambridge},
            postcode={02139}, 
            state={MA},
            country={USA}}
            
\affiliation[inst6]{organization={Department of Civil and Environmental Engineering, Massachusetts Institute of Technology},
            city={Cambridge},
            postcode={02139}, 
            state={MA},
            country={USA}}

\begin{abstract}
The low elastic modulus of soft materials, combined with  geometric nonlinearity and rate dependence, presents significant challenges in the characterization of their mechanical response. We introduce a novel method for measuring the mechanical properties of soft materials under large deformations via cylindrical cavity expansion. In this method, a cylindrical cavity is fabricated in the material and expanded by volume-controlled injection of an incompressible fluid with simultaneous measurement of the applied pressure at the cavity wall. The relationship between applied pressure and deformation at the cavity wall is then employed to characterize the nonlinear mechanical properties. We demonstrate the feasibility of the proposed method and validate it  by measuring  the mechanical properties of synthetic polydimethylsiloxane (PDMS) and comparing with  reported values in the literature. Results indicate that the cylindrical cavitation method effectively captures the response of PDMS over a wide range of stiffness (shear modulus ranging from 5 kPa to 300 kPa) and exhibit high repeatability. The proposed method overcomes limitations in characterization of ultra-soft materials using traditional testing methods, such as challenges with fabrication and clamping in unaxial tension testing and friction and adhesion effects in compression and indentation testing, thus enabling accurate and precise characterization.  It also offers improved accuracy and repeatability over other needle induced cavity expansion methods  due to precise control over the initial cavity dimension and shape at the cost of increased invasiveness of testing.
\end{abstract}




\end{frontmatter}


\section{Introduction}
\label{sec:introduction}
 Soft matter encompasses many materials, such as polymers, foams, gels, colloids, granular materials, and most biological materials. These materials have extensive applications in various fields, including biomedicine~\cite{kong2020fiber,dagdeviren2015conformal}, soft robots~\cite{li2021self}, and  wearable devices~\cite{lee2018stretchable}. Precisely characterizing the mechanical properties of soft materials is fundamental in broadening their application scope. This endeavor enhances the understanding of human organ damage mechanisms among researchers and fosters the advancement of collision protection devices across diverse fields such as sports~\cite{tomin2022polymer}, transportation~\cite{abayazid2024viscoelastic}, and the military~\cite{moss2009skull,gauch2018near}. Ultra-soft materials deform noticeably due to gravity, presenting significant challenges in fabricating and clamping standard specimens~\cite{budday2017mechanical}. Additionally, the surfaces of soft materials often possess adhesive properties, leading to adverse effects from adhesion and friction during indention tests~\cite{mckee2011indentation}. Despite extensive mechanical testing across disciplines using traditional experimental methods such as tension~\cite{estermann2020hyperelastic}, compression~\cite{liu2020highly}, and indentation~\cite{esteki2020new}, on soft materials such as brain tissue and hydrogels, the obtained results exhibit significant variations and low repeatability. Emerging techniques such as elastography~\cite{zhang2023noninvasive,weickenmeier2018magnetic} offer non-contact measurement capabilities, holding substantial promise for biomechanical application. However, such methods are all based on the assumption of linear elasticity, and are incapable of accurately measuring the large deformation mechanical behavior of soft materials. Furthermore, the differential penetration depths and attenuation properties of waves across various frequencies induce a correlation between the outcomes of elastography and the testing frequency, thereby impacting the accuracy of the tests~\cite{guertler2018mechanical}. These methods are also inherently dynamic and thus present additional challenges in isolating the rate independent elastic response from inertial and viscoelastic effects.

An emerging class of testing methods  for minimally invasive mechanical characterization of soft and biological materials employs the expansion of a cavity generated in the material. The cavity can be generated through insertion of a syringe needle such as in Cavitation Rheology (CR)~\cite{zimberlin2007cavitation,crosby2011blowing}  and Volume Controlled Cavity Expansion (VCCE)~\cite{raayai2019volume,chockalingam2021probing} or through ablation by a laser pulse such as in Inertial Microcavitation Rheometry (IMR)~\cite{estrada2018high}.
 In IMR, a laser pulse is used to create a cavity within the material sample, and the subsequent dynamic motion of the cavity is visually tracked and modeled to extract high strain rate material properties~\cite{estrada2018high,yang2020extracting,yang2020strain}. A limitation  of IMR is the requirement of transparent material samples for visual tracking of cavity motion. In needle-based methods, a needle is inserted into the material, and a syringe is used to generate a cavity at its tip, which is then expanded through the injection of pressurized fluid. CR, the pioneering technique in this space, expands the needle induced cavity through a pressure controlled process until a maximum pressure is reached following which the cavity expands unstably. This maximum pressure is assumed to correspond to the
theoretically predicted elastic cavitation instability limit~\cite{gent1959internal}, which is then used to determine the elastic modulus. Despite its successful application to variety of synthetic and biological materials~\cite{zimberlin2010cavitation,zimberlin2010water,cui2011cavitation,delbos2012cavity,chin2013cavitation,blumlein2017mechanical,polio2018cross,fuentes2019using}, the
CR technique often results in fracture of the sample prior to reaching the cavitation instability limit and relies on an a priori assumption on the constitutive response. The VCCE technique remedies these problems by performing a volume controlled expansion of the cavity using injection of an incompressible fluid and then using the pressure–volume data (prior to fracture) to extract the nonlinear elastic and viscoelastic material properties~\cite{raayai2019volume,chockalingam2021probing}. 

The needle based methods suffer from a lack of control over the size and shape of the initially generated cavity/defect, as well as from the generation of cracks and pre-stress from the needle insertion.  This limits the repeatability of testing over multiple probings as well as the accuracy of mechanical characterization since  assumptions of the theoretical formulation are violated in the physical testing (such as spherical symmetry, initially stress free cavity with a well defined radius ). 
In the newly proposed method in this manuscript, we adapt the VCCE method for more accurate, precise and repeatable testing by sacrificing its minimally invasive nature for precise control over the initial cavity size. We do this by  controlled creation of uniform cylindrical cavities with high slenderness ratio during the fabrication of the testing samples, instead of cavity generation through needle insertion. The volume controlled expansion of these cylindrical cavities through fluid injection will instead be used to characterize the mechanical properties of the material. The proposed Cylindrical Volume Controlled Cavity Expansion (C-VCCE) technique will be demonstrated through characterization of synthetic PDMS samples with varying base to cross-linker ratio and comparison with results from uniaxial and conventional VCCE testing.  We show that the C-VCCE method is more accurate and repeatable compared to conventional VCCE while still allowing for testing in the ultrasoft regime (shear modulus $\sim$ 1-50 kPa) where uniaxial tensile testing is not feasible.

This paper is structured as follows: in Section \ref{sec:themethod}, the C-VCCE method is introduced with the theoretical foundations explained in Section \ref{sec:theory}. Next, in Section \ref{sec:experiment}, we explain the experimental setup for our C-VCCE method and the phenomena observed in the tests. Then, in Section \ref{sec:results}, we present the experimental outcomes and conduct a comparative analysis with results obtained from other characterization methods. Lastly, in Section \ref{sec:conclusion}, the key points of this work are summarized.

\section{The C-VCCE Method}
\label{sec:themethod}

\begin{figure*}[h]
    \centering
    \includegraphics[width=180mm]{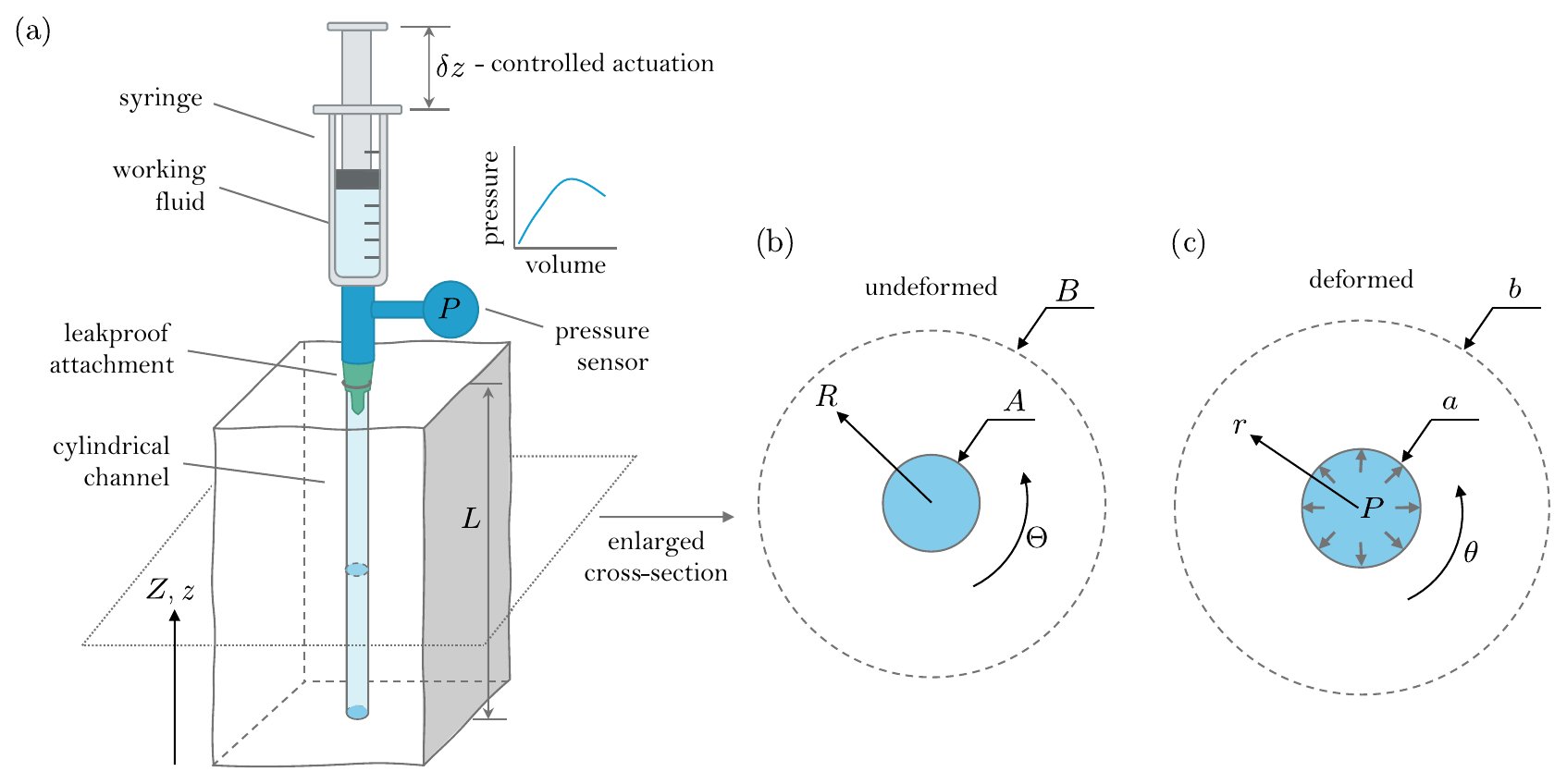}
    \caption{(a) Schematic illustration of the Cylindrical Volume Controlled Cavity Expansion system for characterizing the mechanical response of soft materials, (b) Undeformed state, (c) Deformed state.}
    \label{fig:0}
\end{figure*}

Here we propose the volume controlled expansion of a cylindrical cavity within a material sample along with the concurrent measurement of expansion pressure as a method for accurate and precises measurement of the nonlinear constitutive response of soft materials. 
The precise fabrication of cylindrical cavities is crucial for sample preparation. Here we utilized the casting method with a specific mold to create the cylindrical channel. The prepared sample is sealed at one end to avoid leakage or stress concentrations, connected at the opposite end to a syringe via leak-proof fittings, and fitted with an online pressure sensor (Fig. \ref{fig:0}a with further details in Sec. \ref{sec:experimentsetup}). 
Before sealing the channel, the syringe system is used to fill the system with a degassed incompressible working fluid (typically water). Caution must be exercised in filling the syringe system to avoid any  air bubbles. Once the system is filled and capped, the syringe barrel is depressed via a linear control actuator to inject fluid into the channel at a prescribe volumetric rate while simultaneously recording the pressure. The channel, of undeformed radius $A$, should span the sample, in one direction, but should be located at a radial distance, $B$, from the  boundaries in other directions, such that $B/A>>1$. A more quantitative understanding of this restriction will be provided in the Section \ref{sec:theory}.

\subsection{Volume control and error mitigation}
In this method, accurate and precise control of the injected volume is critical to ensure correct translation of measured data to materials response. Hence, the experimental set-up is designed to minimize compliance and to optimize precision.

Although the need to fabricate a cylindrical cavity within a sample is a complicating factor and might not be straight forward in various materials, this geometry has some significant advantages over the commonly used spherical geometry in VCCE. In spherical cavity expansion, precision of the linear actuation is crucial in minimizing errors for estimation of the cavity radius; for a given precision of the displacement of the plunger, say $\varepsilon z$, the resulting precision in the estimation of the current cavity radius, $a$, is  $\varepsilon a/a=\varepsilon z \cdot S/(4\pi a^3)$, where $S$ denotes the cross-sectional area of the plunger. Hence, the only way to mitigate such errors in the spherical geometry is by reducing $S$ or restricting the experiments to large cavity radii. 
In contrast, for the cylindrical geometry, the precision in estimation of the current cavity radius is  $\varepsilon a/a=\varepsilon z \cdot S/(2\pi a^2 L)$, where $L$ denotes the length of the channel. Accordingly, the length of the  channel can also be tuned to minimize error. 

An additional advantage of long and narrow channels is the mitigation of boundary effects. In the theoretical model, presented next, we will neglect out of plane motion. This assumption holds true away from the ends of the channel. The longer the channel is, the less influence that such boundary effects have on the measurement. Namely, we require $L>>A$.

\subsection{Theoretical formulation}
\label{sec:theory}

Theoretical prediction of the pressure-volume response in the above defined experimental system,  enables comparison with experimental data for identification of constitutive properties.  Provided the slender dimensions of the channel, its volume can be estimated as $V=\pi a^2 L$ and the analysis considers the region of the channel that is unaffected by the boundaries (typically a distance of $\sim 2a$ from the boundary according to Saint-Venant's principle~\citep{love2013treatise}).  
 Accordingly, we consider an infinitely long thick-walled cylinder of undeformed inner and outer radii $(A,B)$, respectively. When subjected to an internal pressure $P$ the cylinder expands in the plane to arrive at the corresponding deformed radii $(a,b)$, as illustrated in Fig. \ref{fig:0}. 
The solution to this problem for incompressible hyperelastic materials  is a classic problem in finite elasticity~\citep{haughton1979bifurcation,hutchinson2020instabilities}. We recapitulate the derivation here for completeness. 
The deformation field is described via a cylindrical coordinate system, such that in the undeformed configuration the body occupies the region
\begin{equation}
    A\leq R \leq B, \qquad 0\leq \Theta < 2\pi, \qquad -\infty < Z < \infty, 
\end{equation}
while in the deformed configuration we have
\begin{equation}
    a\leq r \leq b, \qquad 0\leq \theta < 2\pi, \qquad -\infty < z < \infty, 
\end{equation}

Assuming plane-strain conditions, material points are mapped to the deformed configuration by the relations
\begin{equation}
r=r(R), \qquad \theta=\Theta, \qquad z=Z.
\end{equation}
Restricting our attention to incompressible materials, the remaining unknown function (i.e. $r(R)$) can be found via the volume constraint on any cylindrical sub-region 
\begin{equation}\label{incom}
   r^2-a^2=R^2-A^2 \qquad \to \qquad r(R)=\sqrt{R^2+a^2-A^2}
\end{equation}
The principle stretch components align with the orientation of the coordinates and can be readily written as
\begin{equation}\label{stretches}
    \lambda_r= \frac{\partial r}{\partial R}, \qquad \lambda_\theta=\frac{r}{R}, \qquad \lambda_z=1.
\end{equation}
Incompressibility implies that $\lambda_r\lambda_\theta\lambda_z=1$, which is identically satisfied  by \eqref{incom}. For future use we define \begin{equation}\label{def}
    \lambda=\lambda_\theta=\frac{1}{\lambda_r}=\sqrt{1+\frac{a^2-A^2}{R^2}}, 
\end{equation} where, by using \eqref{incom}, the kinematics of the problem are fully defined for an given set $(A,a)$, as a function of the independent field variable $R$. It remains to determine the corresponding pressure.

The equilibrium requirement in the cylindrical field reduces to a single non-trivial equation 
\begin{equation}
\label{eq:equilib}
   \frac{d\sigma_r}{dr} = \frac{\sigma_\theta-\sigma_r}{r}, 
\end{equation}
where $\sigma_r$ and $\sigma_{\theta}$ are the radial and circumferential principal stress components of the Cauchy stress tensor. Similar to the derivation in~\citep{raayai2019volume}, but for the present cylindrical geometry, we can write the strain energy density in terms of the principal stretches $\hat{W}(\lambda_r,\lambda_\theta, \lambda_z)$.   The principal stress difference is then readily derived as
\begin{equation}
\label{eq:sigthetar}
   \sigma_{\theta}-\sigma_r=\lambda_{\theta}\frac{\partial \hat{W}}{\partial\lambda_{\theta}}-\lambda_r\frac{\partial \hat{W}}{\partial\lambda_{r}}=\lambda W'(\lambda), 
\end{equation}
where $W(\lambda)=\hat{W}(1/\lambda,\lambda,1)$, and the superimposed prime denotes differentiation. 
Substituting the above result in \eqref{eq:equilib} and performing integration  reads
\begin{equation}
\label{eq:Pint}
   P=\int_a^b\lambda W'(\lambda)\frac{\mathrm{d}r}{r}=\int_{\lambda_a}^{\lambda_b}\frac{ W'(\lambda)}{1-\lambda^2}\mathrm{d}\lambda, 
\end{equation}
where we have readily introduced the transformation $\mathrm{d}r/r={\rm d}\lambda/(\lambda(1-\lambda^2))$  as obtained from \eqref{incom} with \eqref{stretches}, used the shorthand notations $(\lambda_a,\lambda_b)=(\lambda|_{r=a},\lambda|_{r=b})$, and applied the  boundary conditions 
\begin{equation}
   \sigma_{r}(r=a)=-P,\qquad \sigma_{r}(r=b)=0.  
\end{equation}

The integral relationship in \eqref{eq:Pint} is the most general result, which applies for any incompressible hyperelastic material. 
The PDMS samples used in this work are well represented by a neo-Hookean model for which 
\begin{equation}
    \hat{W}=\frac{\mu}{2}\left(\lambda_r^2+\lambda_\theta^2+\lambda_z^2-3\right) \qquad \to \qquad W=\frac{\mu}{2}\left(\lambda^2+\frac{1}{\lambda^{2}}-2\right).
\end{equation}
Inserting the above relation in \eqref{eq:Pint} and performing integration, yields 
\begin{equation}\label{eq:innerPab}
    P=\frac{\mu}{2}\left.\left(\lambda^{-2}-2\ln{\lambda}\right)\right|^{\lambda_b}_{\lambda_a}.
\end{equation}
Representative curves obtained using \eqref{eq:innerPab}, for various $B/A$ ratios are shown in Fig. \ref{fig:2 pressure curves}.

Finally, for cylindrical channels that are narrow in comparison with the dimensions of the sample (i.e. for  $B/A\to\infty$), the remote stretch vanishes such that  $\lambda_b\to1$ (see \eqref{def}), and we obtain the useful formula relating the cavity pressure to  cavity stretch in the form
\begin{equation}
\label{eq:innerP}
   P=\frac{\mu}{2}\left(1-\lambda_a^{-2}+2\ln{\lambda_a}\right). 
\end{equation}
An important observation is that this relation predicts that the pressure increases monotonically as the cavity  expands and does not approach an asymptotic limit. Namely, the cavitation instability, a celebrated result in spherical cavity expansion~\cite{gent1959internal}, is absent in the cylindrical filed. However, thick wall tubes are known to exhibit additional types of instability and failure, that can influence the measurement of material properties via channel inflation, as described next.  

\subsection{Mechanical limitations}
Though cavitation is not predicted in the cylindrical geometry, fracture can become a limiting factor. The onset of fracture is highly dependent on the specific material and on imperfections in the sample that can lead to stress-localization. It will be shown in the next section that for the particular case of PDMS fracture is delayed in this geometry, as compared with the spherical cavitation setting~\cite{raayai2019volume,raayai2019capturing} where  imperfections are introduced by the needle insertion process. 

\begin{figure}[h]
    \centering
    \includegraphics[width=90mm]{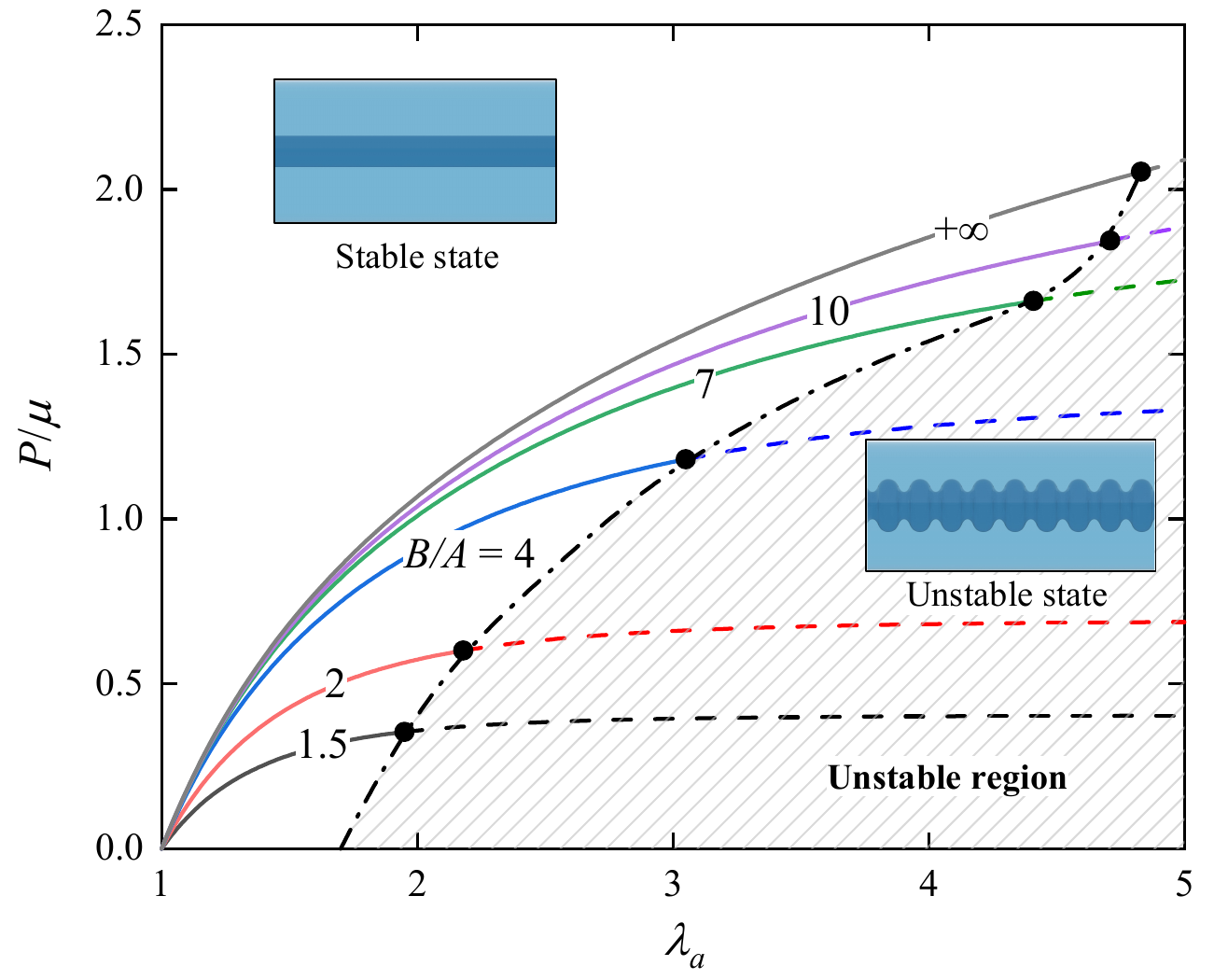}
    \caption{Pressure-stretch curves obtained from (\ref{eq:innerP}) shown for different ratios. In this plot, pressure is normalized by the shear modulus, and the gray shaded area indicates regions in which  the axially symmetric field becomes unstable. Stability limit curve (dash-dotted line) is reproduced from~\cite{cheewaruangroj2019peristaltic}.}
    \label{fig:2 pressure curves}
\end{figure}

An additional limitation is imposed by the potential onset of the `peristaltic instability'  ~\cite{cheewaruangroj2019peristaltic}. 
Beyond a critical pressure, the axially symmetric solution obtained above becomes unstable and the cavity takes an undulating pattern along its length. 
Extending the formulation of Haughton and Ogden~\cite{haughton1979bifurcation} to capture this instability for increasing values of $B/A$, Cheewaruangroj et al.,~\cite{cheewaruangroj2019peristaltic} show that this instability occurs at a finite stretch value.   In Fig. \ref{fig:2 pressure curves}, we use the results of their analysis to  delineate the region of instability in our tests (gray shading). Accordingly, we find the feasibility range of applied stretch (i.e. $\lambda$) in our experiments. We note that for thick cylinders of $B/A\geq10$, instability is triggered at $\lambda>4$, which is beyond the anticipated fracture limit for samples in our experiments.

\subsection{Fitting} 
Once experimental data is obtained, any fitting method can be used to determine the material coefficients for a given constitutive relation \eqref{eq:Pint}. In this work, for the validation in PDMS (Section \ref{sec:experiment}) we will demonstrate the fitting results based on the least squares  method to determine the single material constant - the shear modulus of the material. An important advantage of the the cylindrical geometry is that the radius of the channel (i.e. $A$) is well defined. In contrast, in the VCCE method, the initial defect size is set by the needle insertion process and is thus often included among the fitting parameters~\citep{raayai2019volume,chockalingam2021probing}. 

\section{Validation Experiments in PDMS}
\label{sec:experiment}

Polydimethylsiloxane, often abbreviated as PDMS, is a commonly used silocone rubber. While accessible under brand name  Dow SYLGARD 184,  PDMS has  become ubiquitous in various industries and research labs. It is used for rapid development of microfluidics, flexible electronics, and biomedical devices.   
However, a recent study~\citep{varner2024pdms} reports that its mechanical properties are poorly understood. The most commonly used testing method - tensile testing - is shown to produce elastic moduli that range over 2 orders of magnitude for samples with similar chemical compositions. One explanation for this limitation is that conventional mechanical testing methods are ill suited for highly compliant materials. 


\begin{figure}[!h]
    \centering
    \includegraphics[width=90mm]{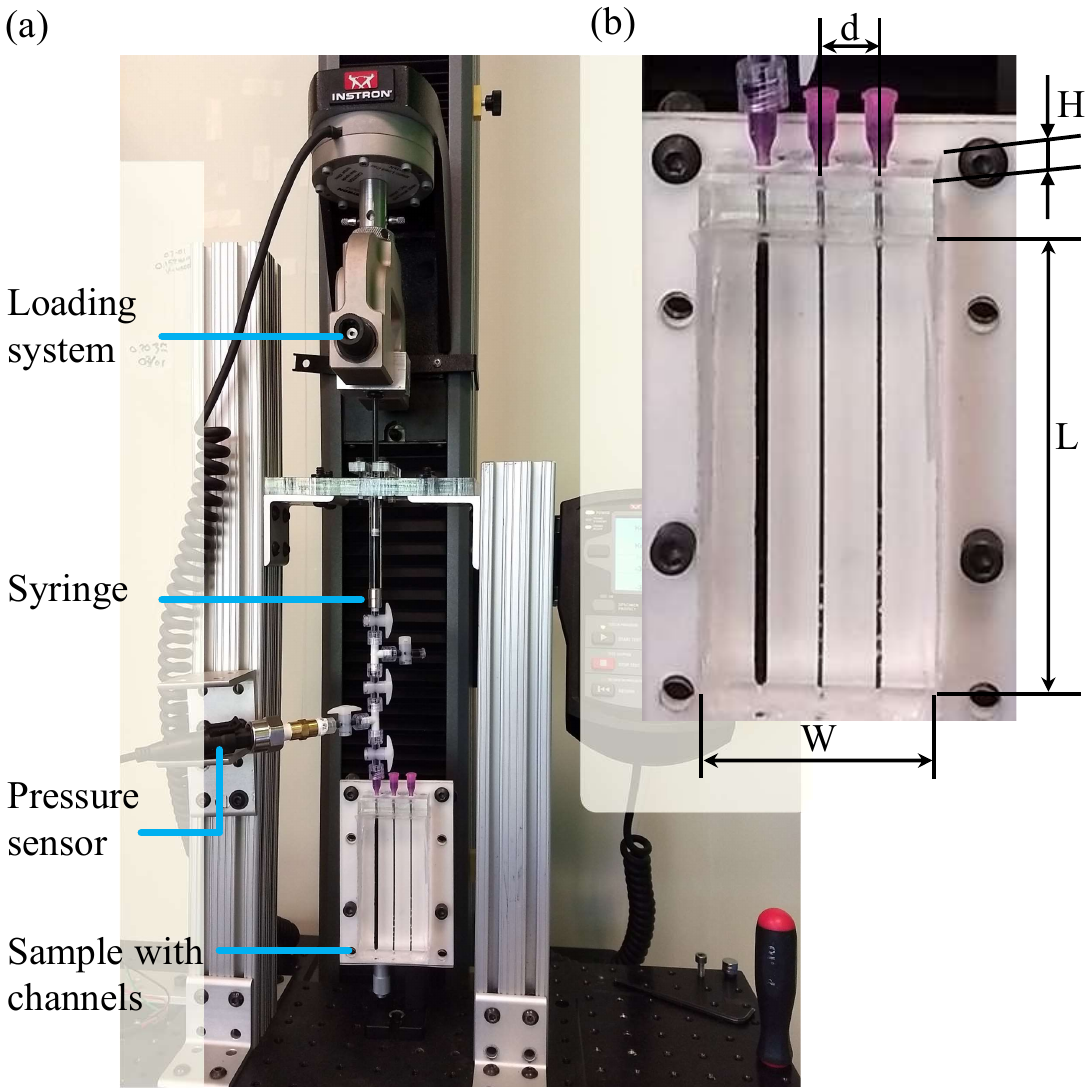}
    \caption{(a) Experimental set up, (b) Sample geometry with \textit{L} = 100 mm, \textit{W} = 70 mm and \textit{H} = 24 mm. Channel spacing \textit{d} is 14 mm on center. }
    \label{fig:3 experimental setup}
\end{figure}

\subsection{Sample preparation}
\label{sec:sampleprep}
In this study, we characterize the elastic properties of  PDMS samples (SYLGARD 184, Dow) with varying base to curing agent ratios (weight:weight) of 15:1, 20:1, 25:1, 30:1, 39.1:1, and 45:1. To ensure thorough mixing of the base and curing agent, we  performed two 1-minute cycles at 2000 RPM in a planetary mixer (THINKY Corporation, USA).
The mixed PDMS was then pored into a rectangular acrylic mold measuring 100 mm in length \textit{L}, 25 mm in height \textit{H}, and 55 mm in width with a channel spacing \textit{d} of 14 mm as shown in Fig. \ref{fig:3 experimental setup}. 
Cylindrical cavities were created in the samples, by inserting three evenly spaced cylindrical carbon fiber rods (radius $A$=0.575 mm) along the length of a rectangular acrylic mold. It should be noted that, to ensure smooth demolding of the sample after solidification, a release agent needs to be sprayed on the surface of the fiber rod. The samples were then degassed inside a vacuum chamber for half an hour  and it was confirmed that there were no visible bubbles inside the samples before transferring them to an oven to cure for 2.5 hours at 100~$^\circ$C. The oven was allowed to cool naturally to room temperature, after which we removed the sample from the mold and extracted the cylindrical rods from within the sample.

\subsection{Experimental setup}
\label{sec:experimentsetup}
The experimental device is shown in Fig. \ref{fig:3 experimental setup}(a). A dedicated frame was constructed on the Instron 5943 universal testing machine (Norwich, USA) to secure the syringe and samples. A gastight  syringe (Hamilton Company, Reno, USA) with a length of 60 \text mm $(h)$ and a volume of 1 mL ($G = S h$) was connected to an online 100PSI $\pm$2\%FS three wire  pressure transducer via a series of luer lock Ts and stopcock valves. Blunt dispensing needles were placed through an acrylic plate and into the cylindrical cavities and secured with cyanoacrylate to create an air tight seal. To ensure accurate measurement of the cavity expansion, the entire system was purged of air and filled with an incompressible working fluid (colored water) prior to initiating the tests.

During an experiment, the Instron depressed the syringe plunger at a constant velocity ($\nu$) of 3~mm/min, injecting fluid and expanding the cylindrical cavity, corresponding to an initial circumferential strain rate at the cavity wall of 0.004~s$^{-1}$ (see Eq.~\eqref{ref:correctrateeq}). The pressure inside the water column was assumed to be uniform throughout;  the measured pressure $(P)$ and  plunger displacement ($\delta z$)  were recorded. 
Given the initial volume of the channel ($\pi A^2 L$), and the change in volume associated with the motion of the plunger ($S\delta z$) we can directly relate the experimental data to the cavity stretch ($\lambda_a$) and stretch rate ($\dot{\lambda_a}$) by
\begin{equation}
   \lambda_a =\sqrt{1+\frac{S\delta z}{\pi A^2 L}}, \qquad  \dot{\lambda_a} = \frac{S \nu}{2 \pi  a A L} \label{ref:correctrateeq}
\end{equation}
 as long as axial symmetry of the channel  is preserved.


\subsection{Observation}
\label{sec:observation}
\begin{figure*}[h]
    \centering
    \includegraphics[width=180mm]{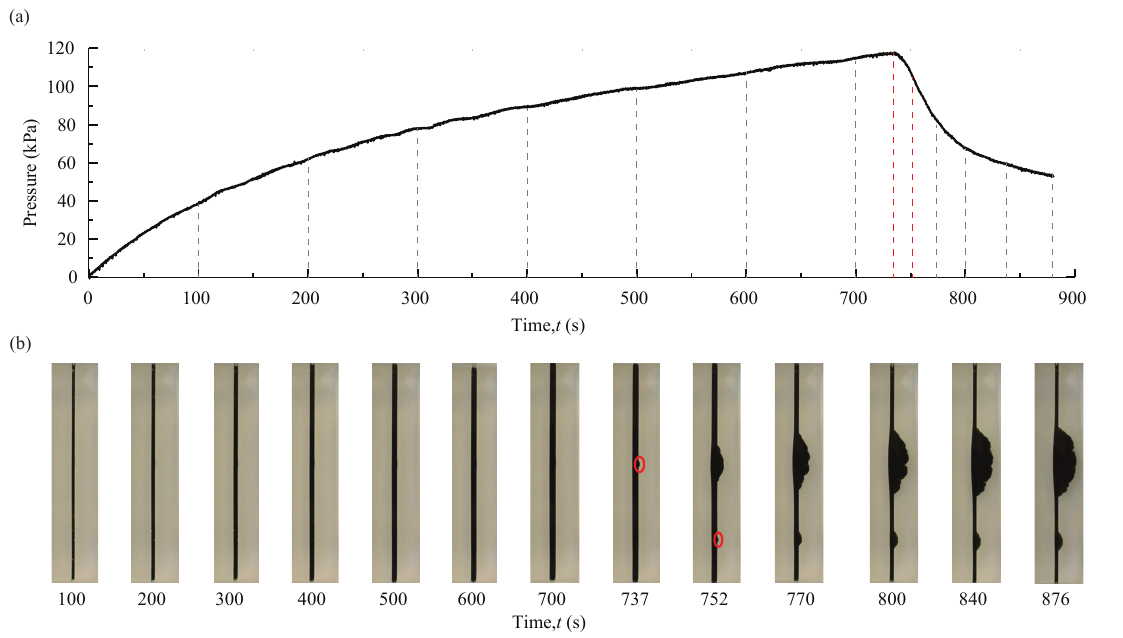}
    \caption{(a) The pressure changes during the process of starting to expand until a fracture occurs, (b) Expansion of cylindrical cavitation at different times.}
    \label{fig:3 experimental results}
\end{figure*}

Fig. \ref{fig:3 experimental results}(a) illustrates the experimental results obtained from a PDMS sample with a curing agent ratio of 25:1, displaying pressure as a function of time. Initially, the cylindrical cavity remains undistorted and filled with liquid, with no pressure change in the cavity. As the fluid is injected, the pressure increases in a monotonic fashion. Upon reaching a value of 117.82 kPa, there is a rapid decline in pressure. The corresponding images in Fig.~\ref{fig:3 experimental results}(b) reveal the uniform circumferential expansion of the cylindrical cavity (black area in the figure) due to the pressure of the injected fluid. At $t$=737 s, the critical pressure threshold is attained, leading to a fracture at the center of the cylindrical cavity (marked by a red circle on Fig. \ref{fig:3 experimental results}(b)) and resulting in a rapid pressure drop. Further fluid injection results in new fractures in the lower portion of the cylindrical cavity at $t$=752 s.

\begin{figure}[h!]
    \centering
    \includegraphics[width=90mm]{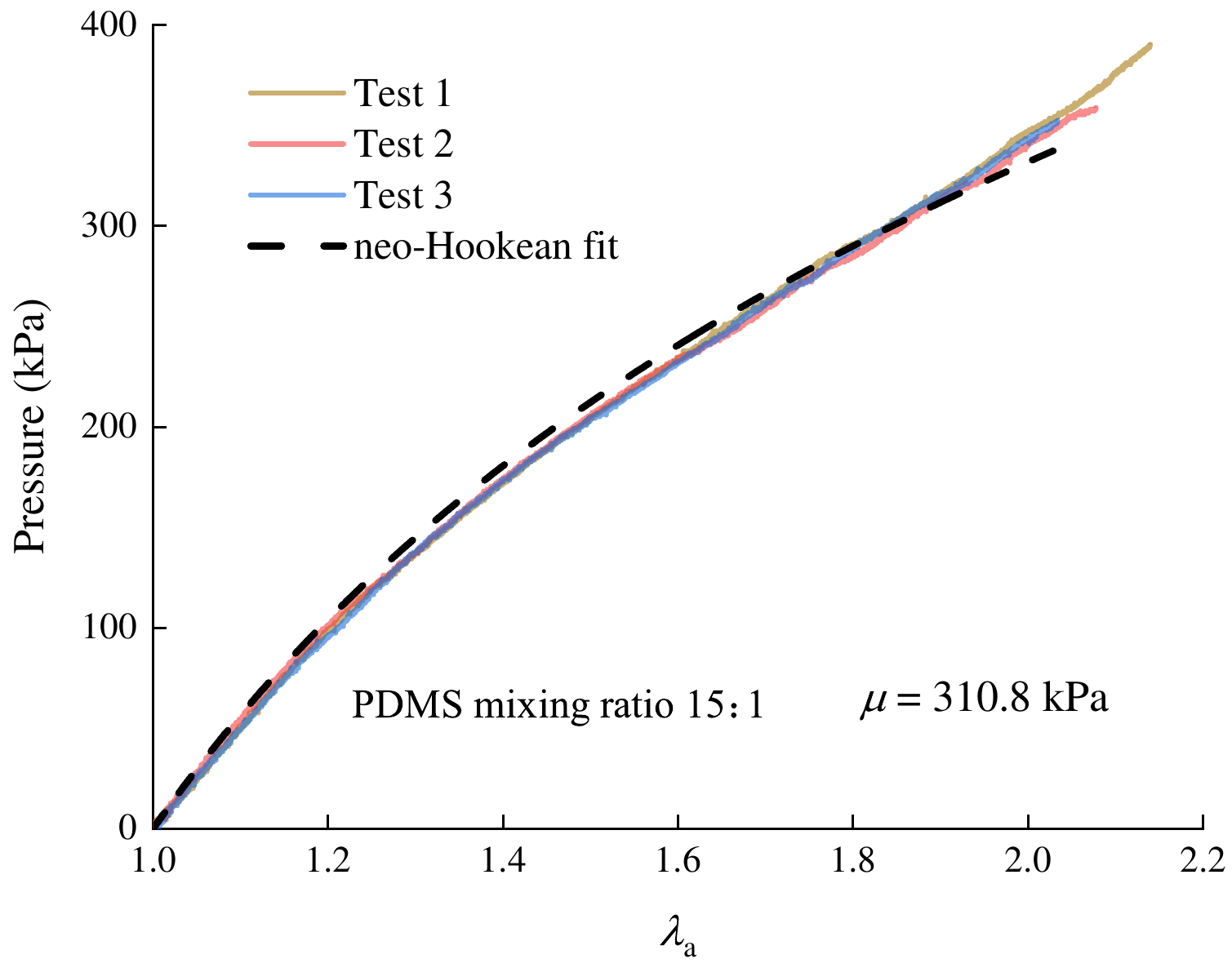}
    \caption{Experimentally measured $P - \lambda_a$ curves for PDMS sample with base to curing ratio of 15:1. The dashed lines correspond to the fitted neo-Hookean response and the fitted shear modulus $\mu$ is indicated on the plot.
    \label{fig:15-1result}}
\end{figure}

\begin{figure*}
    \centering
    \includegraphics[width=180mm]{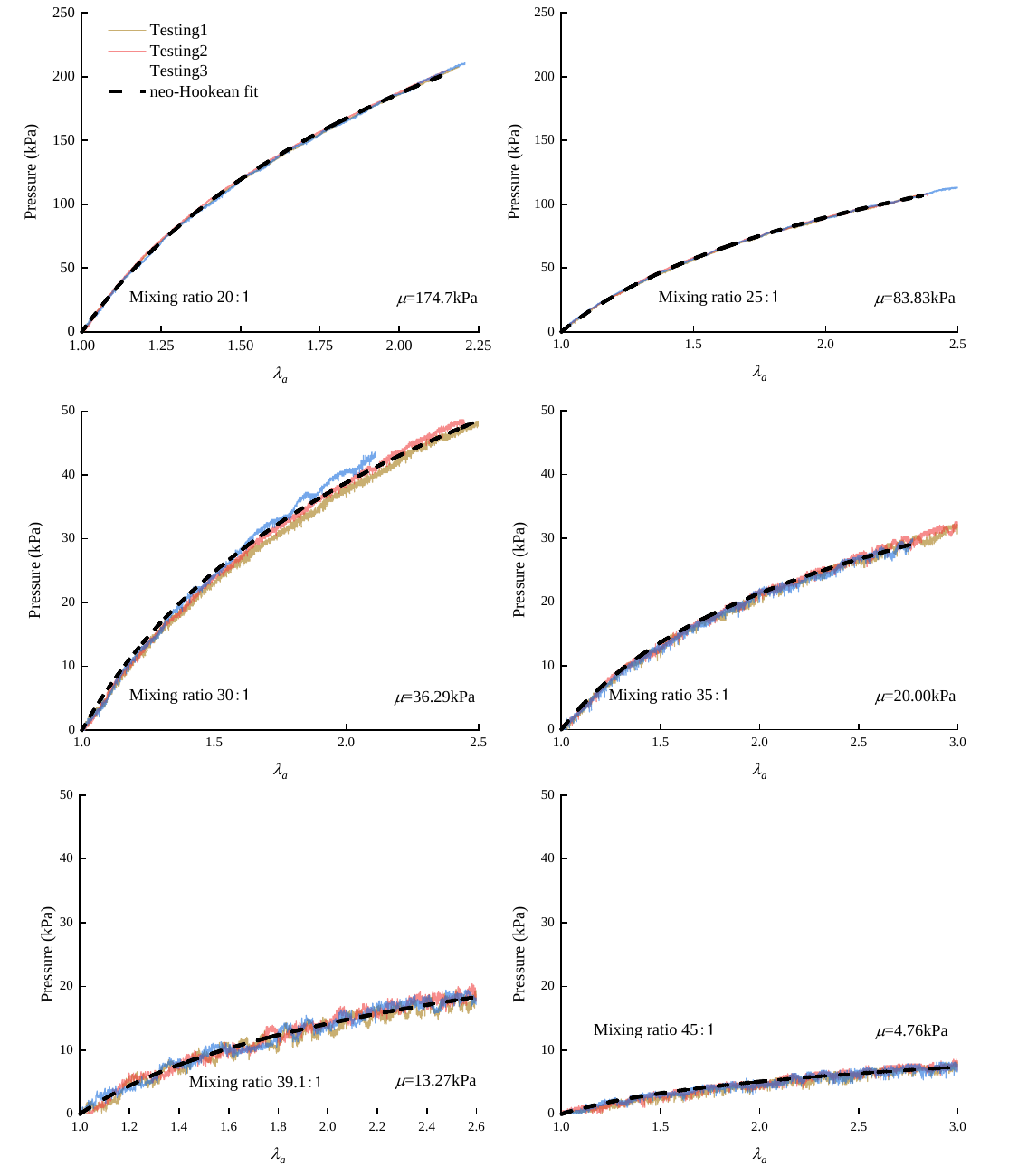}
    \caption{Experimentally measured $P - \lambda_a$ curves for PDMS sample with with varying base to curing agent ratios.  The dashed lines correspond to the neo-Hookean response  fitted by the average of three sets of experimental results and the fitted shear modulus $\mu$ is indicated on the plot. }
    \label{fig:4 seven plots}
\end{figure*}

Given the loading rate of the testing machine and the inner diameter of the syringe, the pressure-time curve can be easily converted into the relationship curve between pressure and stretch. According to Eq.~(\ref{eq:innerP}), the shear modulus of the sample can be fitted, and we will discuss this in detail in the next section.

\section{Results and discussion}
\label{sec:results}

 Fig.~\ref{fig:15-1result} and Fig.~\ref{fig:4 seven plots} illustrate the correlation between fluid pressure within the cavity and radial stretch alongside a fitting curve. We compute the average outcomes of three experiments and fit the material's shear modulus using the least squares method based on Eq.~(\ref{eq:innerP}) The three experiments executed on the same sample exhibit high repeatability, showing strong agreement with theoretical projections. With the incremental rise in the curing agent ratios, the sample exhibit a corresponding softening (evident in the gradual decrease in shear modulus).

It is important to note that the sensor sensitivity remains constant, resulting in more visible oscillation in the experimental data at lower magnitudes. Furthermore, it is apparent from Fig.~\ref{fig:15-1result} and Fig.~\ref{fig:4 seven plots} that both the maximum radial stretch ($\lambda_{max}$) and the critical pressure ($p_{max}$) of the cylindrical cavity are not consistent across experiments performed within the same sample.

Table~\ref{table1} outlines the critical pressure, maximum radial stretch, fitted shear modulus, and correlation coefficients for each test group. Within the three tests performed on each sample, the maximum relative error of the fitted shear modulus is only 2.5\%. The high repeatability of our testing method can be attributed to the regularity and high aspect ratio of the cylindrical cavity shapes. In our test samples, the aspect ratios reach up to 100, significantly reducing the proportion of the non-uniform deformation sections at both ends of the cavity and ensuring that the cavity undergoes uniform circumferential expansion under pressure. These outcomes illustrate our characterization method's high repeatability and the precision of results.

To verify the accuracy of the fitting results, we list the $N^2$ of each set of fitting results in Table~\ref{table1}. $N^2$ is coefficient of determination, defined as $N^2 = SS_R/SS_T$, where $SS_R$ (Sum of Squares Residuals) represents the sum of squared deviations between the predicted values from the model and the actual observed values, and $SS_T$ (Total Sum of Squares) is the sum of squared deviations between the observed values and the mean of the dependent variable. Note that the better fitting, the $N^2$ is closer to 1. As shown in Table~\ref{table1}, the $N^2$  for each set surpasses 0.9, signifying a high level of reliability in our fitting outcomes.

For samples with a mixing ratio of 20:1, the critical pressure  and the maximum radial stretch  from the three trials exhibit noted deviations, with the maximum relative errors being 36.2\% and 28.6\%, respectively. For the remaining samples, the $p_{max}$ and the $\lambda_{max}$, obtained across three trials demonstrate a relative error of approximately 10\%. We attribute this discrepancy primarily to the process of removing the cylindrical rod  from the mold. This action induces unavoidable damage to the interior wall of the cavity, resulting in variations to the critical pressure required for fracture initiation.

\begin{figure}[h!]
    \centering
    \includegraphics[width=90mm]{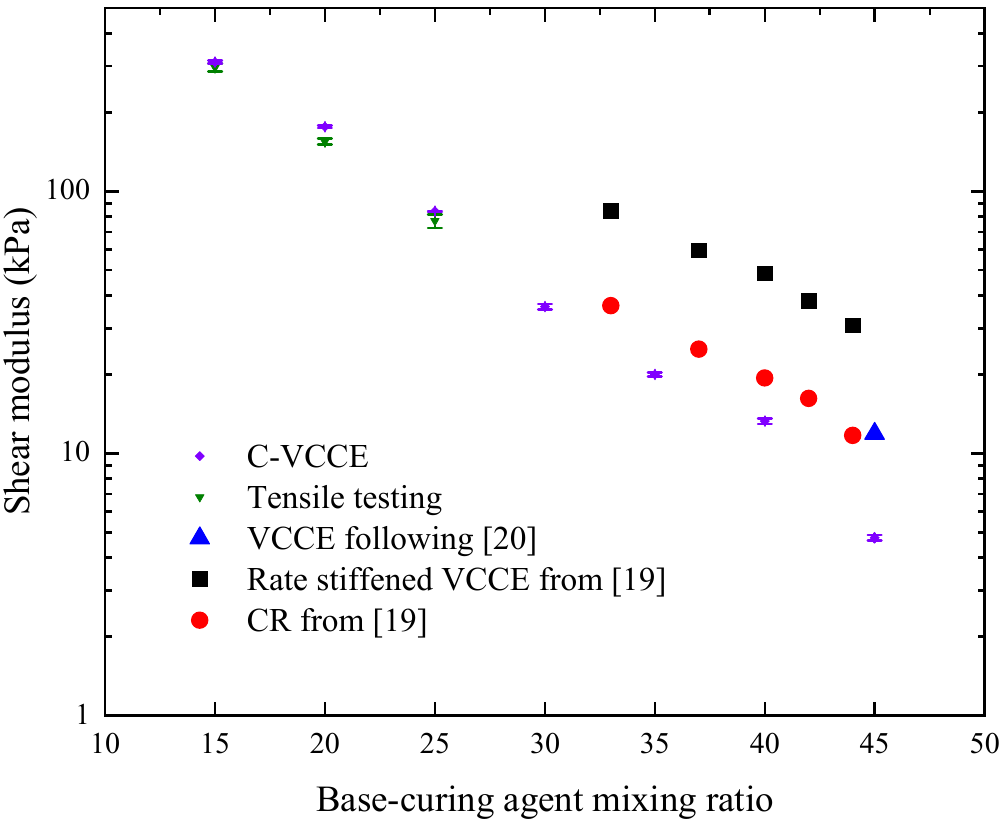}
    \caption{Comparison between the shear moduli of PDMS samples measured using the C-VCCE method proposed in this paper and the test results calculated by the conventional CR method and the VCCE method reported by Shabnam and coworkers~\cite{raayai2019volume}.}
    \label{fig:5 compare}
\end{figure}

To verify the accuracy of the newly introduced characterization method, we prepared additional samples with crosslinking agent ratios of 15:1, 20:1, and 25:1, specifically for quasi-static tensile testing (since tensile testing at higher mixing ratios is not physically feasible due to soft, sticky, and highly deforming samples). The comparison of the shear modulus obtained via different testing methods is shown Fig. \ref{fig:5 compare}. First, we see that the tensile testing results generally agree well with that of the C-VCCE method but nevertheless result in consistently softer moduli estimates – specifically a relative reduction in shear modulus of 6.8\%, 12.1\%, and 8.0\% for mixing ratios of 15:1, 20:1 and 25:1 respectively. This discrepancy could arise from errors in the C-VCCE testing such as from damage to the inner cavity wall during carbon fiber rod extraction, or from errors in the tensile testing of soft samples due to difficulties with fabrication, clamping and control over deformation. In contrast to the good agreement of C-VCCE estimates with that of tensile testing, the shear moduli reported by VCCE testing in \cite{raayai2019volume} are seen to be significantly stiffer. A major reason for the discrepancy is the fact that the constant volumetric rate testing in \cite{raayai2019volume} leads to high stretch rates during initial cavity expansion (maximum rate $\sim $ 1-10 s$^{-1}$)  that results in the probing/measurement of a rate stiffened instantaneous modulus (see Sec 3.7, 6.3 of \cite{chockalingam2021probing}). The shear modulus estimated from an improved version of the VCCE method reported in \cite{chockalingam2021probing} that isolates the rate independent shear modulus from the viscoelastic properties is shown in Fig. \ref{fig:5 compare} for the case of mixing ratio of 45:1. This improved estimate compares more favorably with the C-VCCE value but is nevertheless still noticeably stiffer. This residual discrepancy is very likely arising from the lack of control over the size and shape of the cavity in the VCCE method and the C-VCCE results are likely closer to the ground truth. This also reiterates the need for additional reliable testing methods for extremely soft materials. Estimates from the Cavitation Rheology method are also seen to be stiffer compared to C-VCCE results, the discrepancy likely arising from the various inherent limitations of the technique mentioned in the Introduction.
 
Although the C-VCCE method allows for reliable testing of much softer materials compared to uniaxial testing, it has certain limitations. In uniaxial tension, the gauge length of the specimen undergoes uniform deformation, allowing the stress-strain relationship obtained from the test to directly reflect the intrinsic properties of the material. In contrast, in our method, while the cylindrical cavity can be considered as uniformly expanding, the  strain distribution across the entire configuration is not uniform but radially decays. This results in varying strain rates at different locations. This prohibits rate dependent testing at a fixed strain/stretch rate. Nevertheless, the maximum stretch rates (which will be at at the cavity wall and at the initial cavity size) can be controlled, and thus quasistatic testing is still possible such as in this manuscript where it was restricted to $\sim$ 10$^{-3}$ s$^{-1}$. Additionally, while direct extraction of the stress-strain curves at different strain rates may not be feasible, one can still come up with ways to extract the rate dependent material properties, following a similar approach as in \cite{chockalingam2021probing}. This will follow in future work.

\section{Conclusion}
\label{sec:conclusion}
In this work, we present a novel method for characterizing the mechanical properties of soft materials using cylindrical cavity expansion (C-VCCE). Based on the theoretical curve of cylindrical cavity expansion, we obtain the shear modulus of PDMS samples with different curing agent ratios through fitting experimental results. This method exhibited high accuracy and repeatability due to the uniform deformation of cylindrical cavities with regular shapes and long lengths. Future work could consider improvements based on this method to characterize the mechanical behavior of materials at medium strain rates.

The test results of this approach exhibit high consistency compared to traditional tensile methods and are capable of characterizing softer materials. Unlike conventional CR techniques, this method does not rely on critical pressure, thus showing insensitivity to initial defect sizes. Given that this method relies on fabrication of  cylindrical cavities, it is more suited for characterizing synthetic soft materials rather than biological materials.

\begin{table*}
\begin{center}
\begin{tabular}{ c c c c c } 
\hline
Curing agent ratio & $\mu$ (kPa) & $N^2$ & $p_{max}$ (kPa) & $\lambda_{max}$\\
\hline
15:1-1	&  308.92&  0.99&  	390.44&  	2.14\\
15:1-2	&  315.11&  0.99&  	358.92&  	2.08\\
15:1-3	&  308.53&  0.99&  	352.76&  	2.04\\
\hdashline
20:1-1	&  178.52& 	0.99&  	161.95&  	1.78\\
20:1-2	&  175.61& 	0.99&  	221.93&  	2.29\\
20:1-3	&  174.42& 	0.99&  	210.21&  2.21\\
\hdashline
25:1-1	&  83.41&  	0.99&  	107.22&  	2.36\\
25:1-2	&  83.92&  	0.99&  	108.55&     2.39\\
25:1-3	&  84.15&  	0.99&  	117.82&  	2.58\\
\hdashline
30:1-1	&  35.40&  	0.99&  	49.91&  	2.46\\
30:1-2	&  36.32&  	0.99&  	49.14&  	2.51\\
30:1-3	&  37.15&  	0.99&  	40.94&  	2.11\\
\hdashline
35:1-1	&  19.94&  	0.99&  	33.18&  	3.12\\
35:1-2	&  20.36&  	0.99&  	34.58&  	3.30\\
35:1-3	&  19.69&  	0.99&  	29.95&  	2.79\\
\hdashline
39.1:1-1&  	12.91& 	0.93&  	21.43&  	2.65\\
39.1:1-2&  	13.57& 	0.95&  	23.74&  	2.65\\
39.1:1-3&  	13.34& 	0.94&  	21.12&  	2.65\\
\hdashline
45:1-1	&  4.66&  	0.95&  	7.89&  	3.21\\
45:1-2	&  4.89&  	0.96&  	7.84&  	3.26\\
45:1-3	&  4.74&  	0.96&  	7.86&  	3.26\\
\hline
\end{tabular}
\caption{Parameters for fitting the shear modulus of samples with different base–cross-linker ratios based on the experimentally measured pressure and using Eq. \ref{eq:innerP}.The agent ration suffix number represents the number of experiments. The agent ratio suffix number (-1, -2, -3) represents the testing number.}
\label{table1}
\end{center}
\end{table*}

\section{Acknowledgements}
J.L. acknowledges the support of the National Key R\&D Program of China (2022YFB4300101).

\appendix
\section{Tension testing}
\label{sec:sample:appendix}
Here, we present the quasi-static tensile test (strain rate 0.005~s$^{-1}$) results of samples with three different curing agent ratios. The sample size is consistent with the size of the F-type sample in the Standard Test Methods~\cite{american2006standard}. Compared to the cylindrical cavitation method, the tensile test shows significantly lower repeatability, particularly when the material has a low shear modulus. Across the three experimental sets, the maximum error observed in the shear modulus amounts to 8.7\%.

\begin{figure*}[h!]
    \centering
    \includegraphics[width=\textwidth]{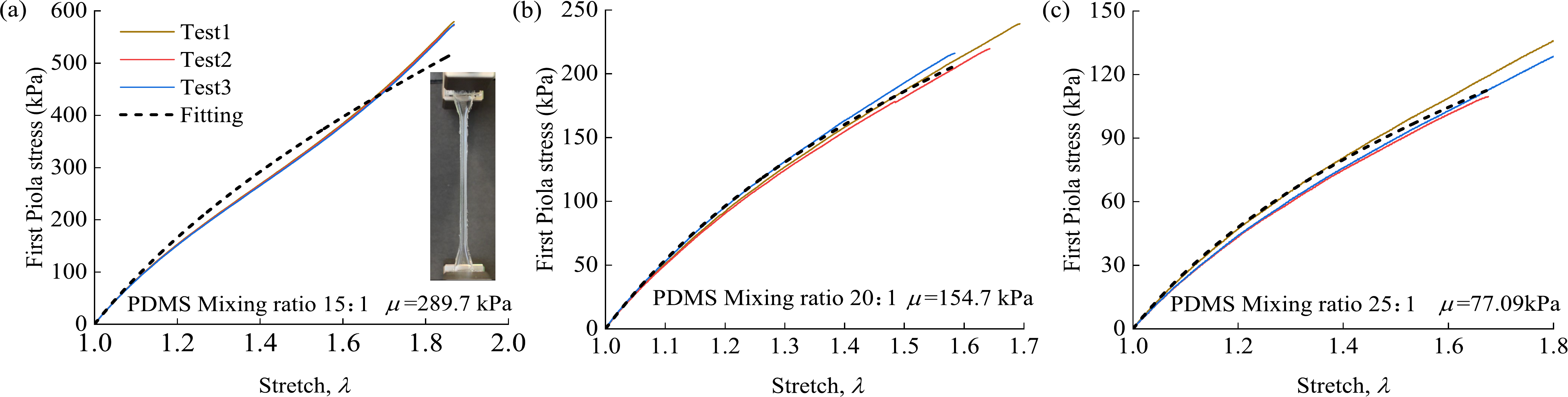}
    \caption{Plots of stress-longitudinal stretch ($\lambda$) curves for uniaxial testing of PDMS samples with three different base to cross-linker ratios: (a) 15:1, (b) 20:1, (c) 25:1.}
    \label{fig:A1 Tension}
\end{figure*}

 \bibliographystyle{elsarticle-num} 
 \bibliography{bib}





\end{document}